\documentstyle[pre,aps,twocolumn,psfig]{revtex}
\begin{document}
\wideabs{
\title {Thermodynamics of dissipative quantum systems by effective potential}
\author{Alessandro Cuccoli\cite{email}, Andrea Rossi\cite{email},
        Valerio Tognetti\cite{email}}
\address{Dipartimento di Fisica dell'Universit\`a di Firenze and
         Istituto Nazionale di Fisica della Materia - INFM,\\
         largo Enrico Fermi~2, I-50125 Firenze, Italy}
\author{Ruggero Vaia\cite{email}}
\address{Istituto di Elettronica Quantistica
         del Consiglio Nazionale delle Ricerche,
         via Panciatichi~56/30, I-50127 Firenze, Italy}
\date{Phys. Rev. E {\bf 55}, R4849 (1997)}
\maketitle
\begin{abstract}
Classical-like formulas are given in order to evaluate thermal averages of
observables belonging to a quantum nonlinear system with dissipation
described by the Caldeira-Leggett model [Phys. Rev. Lett. {\bf 46}, 211
(1981); Ann. Phys. (N.Y.) {\bf 149}, 374 (1983)\protect{]}. The underlying
scheme is the {\it pure-quantum self-consistent harmonic approximation},
which leads to expressions with a Boltzmann factor involving an {\it
effective potential} and with a Gaussian average. The latter describes the
effect of the fluctuations of purely quantum origin.
As an illustration we calculate the coordinate probability distribution for
a double-well potential in the presence of various degrees of Ohmic
dissipation.
\end{abstract}
%\pacs{05.30.-d, 05.40.+j, 05.70.-a}
% 05.20.-y  Statistical mechanics
% 05.30.-d  Quantum statistical mechanics
% 05.40.+j  Fluctuation phenomena, random processes, and Brownian motion
% 05.70.-a  Thermodynamics
}

%ßßßßßß Paper text ßßßßßßßßßßßßßßßßßßßßßßßßßßßßßßßßßßßßßßßßßßßßßßßßßß

The concept of effective potential, meant to reduce quantum statistical
mechanics calculations to classical ones was firs introduced by
Feynman\cite{Feynman55,Feynman72}. He introduced a variational principle
for the path-integral expression of the partition function --- the {\it
Feynman-Jensen} (FJ) {\it inequality} --- and used it with a ``free
particle'' trial action.

A significant improvement has been achieved by Giachetti and Tognetti
\cite{GT85prl,GT86prb} and independently by Feynman and Kleinert
\cite{FeynmanK86} using a quadratic trial action with the same variational
principle. For nonstandard systems, where the FJ inequality is generally
not valid, the {\it pure-quantum self-consistent harmonic approximation}
(PQSCHA) gives a way to construct an effective Hamiltonian, thus recovering
the phase-space concept and the classical-like formulas for thermal
averages \cite{CTVV92ham,CGTVV95}.

Several successful applications in different branches of condensed matter
physics pointed out the power of the approach \cite{CGTVV95}. In the case
of open systems, little work has been done for taking into account the
quantum dissipation in the effective potential formalism; indeed, in a {\it
system-plus-reservoir} model, only the expression of the effective
potential for expressing the partition function as a configuration integral
has been given, both for linear \cite{Falci91,Weiss93} and nonlinear
coupling \cite{BaoZW95} with environmental oscillators.

By using the PQSCHA approach, that is equivalent to the variational method,
we present here an accurate derivation of the density matrix of a nonlinear
system interacting with a heat bath through the {\it Caldeira-Leggett} (CL)
{\it model} \cite{CaldeiraL}. This model considers the system of interest
as linearly interacting with a bath of harmonic oscillators, whose
coordinates can be integrated out from the path integral, leaving the CL
Euclidean action
\begin{eqnarray}
 S[q(u)] &=&\int_0^{\beta\hbar}{du\over\hbar}
 \left[ {m\over2}\dot q^2(u) + V\big(q(u)\big) \right]
\nonumber\\
 &-& \int_0^{\beta\hbar}{du\over4\hbar}
 \int_0^{\beta\hbar}\!\! du'\,k(u{-}u')\,\big(q(u)-q(u')\big)^2~.
 \label{e.S}
\end{eqnarray}
The kernel $k(u)=k(\beta\hbar-u)$ is a function that depends on temperature
$T=1/\beta$ and on the spectral density of the environmental bath
\cite{Weiss93}. The density matrix elements in the coordinate
representation are expressed by Feynman's path integral as
\begin{equation}
 \rho(q'',q')\equiv\langle q''|\hat\rho|q'\rangle
 =\int_{q'}^{q''}{\cal{D}}[q(u)]~e^{-S[q(u)]}~,
\label{e.rhoqq}
\end{equation}
where the path integration is defined as a sum over all paths $q(u)$,
with $u\in[0,\beta\hbar]$, $q(0)=q'$, and $q(\beta\hbar)=q''$.

As suggested by Feynman \cite{Feynman72}, we can rearrange the path
integral (\ref{e.S}) summing over classes of paths that share the same
average point
\begin{eqnarray}
 \rho(q'',q')&=&\int d\bar{q}~ \bar\rho(q'',q';\bar{q})~,
\label{e.rhobrho}
\\
 \bar\rho(q'',q';\bar{q}) &=& \int_{q'}^{q''} \!\!\! {\cal{D}}[q(u)]\,
 \delta\big(\bar q-\bar q [q(u)]\big)
 ~e^{-S[q(u)]}~,
\label{e.brho}
\end{eqnarray}
where $\bar{q}[q(u)]=(\beta\hbar)^{-1}\int_0^{\beta\hbar}du\,q(u)$ is the
average-point functional.
Furthermore, as only paths with a fixed average point $\bar{q}$ 
appear into the path integral~(\ref{e.brho}),
from action~(\ref{e.S}) we define a trial action $S_0[q(u)]$
by replacing $V\big(q(u)\big)$ with a trial quadratic
``potential''
\begin{equation}
 V_0(q;\bar{q}) = w(\bar{q})
 +{\textstyle{1\over2}} m\omega^2(\bar{q})\,(q-\bar{q})^2~,
\label{e.V0}
\end{equation}
where the parameters $w(\bar{q})$ and $\omega^2(\bar{q})$ are now to be
optimized, so that the trial reduced density $\bar\rho_0(q,q;\bar{q})$ at
best approximates $\bar\rho(q,q;\bar{q})$, {\it for each value of}
$\bar{q}$. Note that $V_0(q;\bar{q})$ is not related to a quantum
observable, since it depends on $\bar{q}$: so, the first part of the trial
action is also nonlocal.

The evaluation of the trial reduced density $\bar\rho_0(q'',q';\bar{q})$,
in spite of the fact that the action is quadratic, is rather tricky: on the
one hand, the known general method of calculating the minimal action fails
since its minimization gives rise to infinite order equations of motion,
and, on the other hand, the method of Fourier expansion of the paths in
terms of discrete Matsubara components conflicts with the openness of the
paths, since we are indeed looking also for the off-diagonal part of the
density matrix ($q''\neq{q'}$). However, the second way may still be
followed by transforming as
\begin{eqnarray}
 \bar\rho(q'',q';\bar{q}) &=&
\nonumber\\
 & & \hspace{-15mm}
 \lim\limits_{\varepsilon\to 0}
 \bigg\{ {1\over{\cal F}_\varepsilon}
 \oint{\cal{D}}[q(u)] ~\delta\Big(\bar q-\bar q [q(u)]\Big)
 ~e^{-S[q(u)]}\bigg\}~;
\label{e.brho0eps}
\end{eqnarray}
the integral is over all {\it closed} paths
$\{q(u)\,|\,u\in[0,\beta\hbar]\,\}$ that satisfy the constraints
$q(\varepsilon)=q'$ and $q(\beta\hbar-\varepsilon)=q''$.
${\cal{F}}_\varepsilon$ is the integral over the open paths
$\{\,q(u)\,|\,u\in[-\varepsilon,\varepsilon]\,\}$ (the range
$[\beta\hbar-\varepsilon,\beta\hbar]$ is periodically mapped onto
$[-\varepsilon,0]$) with end points $q(-\varepsilon)=q''$ and
$q(\varepsilon)=q'$; for small $\varepsilon$ it is dominated by the kinetic
contribution
\begin{equation}
 {\cal F}_\varepsilon = \sqrt{m\over4\pi\hbar^2\varepsilon}
 ~\exp\left[-{m\over4\hbar\varepsilon}(q''{-}q')^2+O(\varepsilon)\right]~.
\end{equation}
Now the paths in Eq.~(\ref{e.brho0eps}) can be Fourier expanded,
\begin{equation}
 q(u)=\bar{q}+2\sum\limits_{n=1}^{\infty}~(x_n~\cos\nu_nu+y_n~\sin\nu_nu)~;
\label{e.qn}
\end{equation}
$\nu_n=2\pi{n}/(\beta\hbar)$ are the Matsubara frequencies and the
$\nu_0$ component is just $\bar{q}$, as it appears from the inverse
transformation. The measure of the path integral then becomes \cite{Feynman72}
\begin{equation}
 \sqrt{m\over2\pi\hbar^2\beta}\,\int\,d\bar{q}~\prod\limits_{n=1}^{\infty}
 {m\beta\nu_n^2\over\pi}\,\int dx_n\,dy_n~,
\end{equation}
and the trial action takes then the form
\begin{eqnarray}
 S_0[q(u)] &=& \beta w(\bar{q})
\nonumber\\
 & & \hspace{-5mm} +\beta m \sum\limits_{n=1}^{\infty}
 \big[\nu_n^2+\omega^2(\bar{q})+\nu_n\gamma(\nu_n)\big]
 (x_n^2+y_n^2).
\end{eqnarray}
Here we have made use of the relation that connects the kernel $k(u)$ with
the Laplace transform $\gamma(z)$ of the real-time memory
damping function $\gamma(t)$ \cite{Weiss93}
\begin{equation}
 k(u) = {m\over\beta\hbar} \sum\limits_{n=-\infty}^{\infty}
 e^{i\nu_n u}~|\nu_n|~\gamma\big(z=|\nu_n|\big)~.
\end{equation}
While it is trivial to manage the $\delta$ function that fixes $\bar{q}$, the
end-point constraints are implemented by inserting the $\delta$ functions
$\delta\big(q(\beta\hbar-\varepsilon)-q'')$
and $\delta\big(q(\varepsilon)-q')$, and
then using their Fourier representation. The calculation of $\bar\rho_0$
can then be carried forward by Gaussian quadratures giving
\begin{eqnarray}
 \bar\rho_0(q'',q';\bar{q}) &=& \sqrt{m\over2\pi\hbar^2\beta}
 ~{e^{-\beta w(\bar{q})}\over\mu(\bar{q})}
\nonumber\\
 & &\times\lim\limits_{\varepsilon\to 0} \Bigg[
 {1\over{\cal F}_\varepsilon}
 {e^{-\xi^2/c_\varepsilon} \over\sqrt{\pi c_\varepsilon}}\,
 {e^{-\zeta^2/s_\varepsilon} \over\sqrt{\pi s_\varepsilon}}
 \Bigg]~,
\label{e.brho0eps1}
\end{eqnarray}
where $\xi\equiv{\textstyle{1\over2}}(q'+q'')-\bar{q}$
and $\zeta\equiv{q''-q'}$,
\begin{eqnarray}
 \mu(\bar{q})&=&\prod\limits_{n=1}^{\infty}
 {\nu_n^2+\omega^2(\bar{q})+\nu_n\gamma(\nu_n) \over \nu_n^2}~,
\label{e.mu}
\\
 c_\varepsilon &=& {4\over\beta m}~ \sum\limits_{n=1}^{\infty}
 {\cos^2\nu_n\varepsilon\over\nu_n^2+\omega^2(\bar{q})+\nu_n\gamma(\nu_n)}~,
\label{e.ceps}
\\
 s_\varepsilon &=& {16\over\beta m}\, \sum\limits_{n=1}^{\infty}
 {\sin^2\nu_n\varepsilon\over\nu_n^2+\omega^2(\bar{q})+\nu_n\gamma(\nu_n)}~.
\label{e.seps}
\end{eqnarray}
In the last expressions, the leading terms for small $\varepsilon$ are
$c_\varepsilon=2\alpha(\bar{q})+O(\varepsilon)$~, and, in a less
straightforward way \cite{CaldeiraL,Kirsten},
$s_\varepsilon=4\hbar\varepsilon/m
\big[1-{2\varepsilon\over\hbar m}\lambda(\bar{q})+o(\varepsilon)\big]$,
with
\begin{eqnarray}
 \alpha(\bar{q}) &=& {2\over\beta m}~ \sum\limits_{n=1}^{\infty}
 {1\over \nu_n^2+\omega^2(\bar{q})+\nu_n\gamma(\nu_n)}~,
\label{e.alpha}
\\
 \lambda(\bar{q}) &=& {m\over\beta}\,\sum\limits_{n=-\infty}^{\infty}
 {\omega^2(\bar{q})+|\nu_n|\gamma(|\nu_n|)
 \over \nu_n^2+\omega^2(\bar{q})+|\nu_n|\gamma(|\nu_n|)}~.
\label{e.lambda}
\end{eqnarray}
Eventually, the limit in $\varepsilon$ can be easily taken, and the result
reads
\begin{equation}
 \bar\rho_0(q'',q';\bar{q}) = \sqrt{m\over2\pi\hbar^2\beta}
 ~{e^{-\beta w(\bar{q})}\over\mu(\bar{q})}
 ~{e^{-\xi^2/2\alpha(\bar{q})-\lambda(\bar{q})\zeta^2/2\hbar^2}
 \over\sqrt{2\pi \alpha(\bar{q})} }~.
\label{e.brho0}
\end{equation}
Now, if $\rho_0(q'',q')$ -- obtained using the last result in
Eq.~(\ref{e.rhobrho}) -- is taken as an approximation for the exact
density matrix $\rho(q'',q')$, we have a way to explicitly calculate any
quantum thermal average by two Gaussian quadratures plus a single integral
in $\bar{q}$. A more convenient formalism deals with phase space: let us
briefly show it.

The Weyl symbol for an observable
$\hat{\cal{O}}=\hat{\cal{O}}(\hat{p},\hat{q})$ is the phase-space function
${\cal{O}}(p,q)$ that is defined in terms of the matrix elements in
coordinate space as \cite{Berezin80}
\begin{equation}
 {\cal{O}}(p,q)=\int d\zeta~e^{-ip\zeta/\hbar}
 \big\langle{q{+}{\textstyle{1\over2}}\zeta\big|
 \hat{\cal{O}}\big|q{-}{\textstyle{1\over2}}\zeta}\big\rangle~.
\end{equation}
By means of a simple property of Weyl symbols the average of $\hat{\cal{O}}$
takes the form of a phase-space integral,
\begin{equation}
 \big\langle{\hat{\cal{O}}}\big\rangle =
 {1\over{\cal{Z}}} \int {dp\,dq\over2\pi\hbar}~\rho(p,q)~{\cal{O}}(p,q)~.
\label{e.aveWeyl}
\end{equation}
From Eqs.~(\ref{e.rhobrho}) and~(\ref{e.brho0}) the
Weyl symbol for the trial density operator turns out to be
\begin{eqnarray}
 \rho_0(p,q) &=& 2\pi\hbar \sqrt{m\over2\pi\hbar^2\beta}
 \int d\bar q\,{e^{-\beta w(\bar{q})}\over\mu(\bar{q})}
\nonumber\\
 & & \hspace{20mm}
 \times {e^{-\xi^2/2\alpha(\bar{q})} \over\sqrt{2\pi\alpha(\bar{q})}}
 \,{e^{-p^2/2\lambda(\bar{q})} \over\sqrt{2\pi\lambda(\bar{q})}}~,
\label{e.rho0pq}
\end{eqnarray}
where $\xi\equiv{q}-\bar{q}$. Therefore, the average of any observable
$\hat{\cal{O}}(\hat{p},\hat{q})$ can be expressed as a classical formula
\begin{equation}
 \big\langle{\hat{\cal{O}}}\big\rangle
 ={1\over{\cal{Z}}} \sqrt{m\over{2\pi\hbar^2\beta}}
 \int\,d\bar{q} ~\big\langle\!\!\big\langle {{\cal{O}}(p,\bar{q}+\xi)}
 \big\rangle\!\!\big\rangle
 ~e^{-\beta V_{\rm{eff}}(\bar{q})}~,
\label{e.avepq}
\end{equation}
where the double bracket is the Gaussian average operating over $p$ and
$\xi$, with moments $\langle\!\langle\xi^2\rangle\!\rangle=\alpha(\bar{q})$
and $\langle\!\langle{p^2}\rangle\!\rangle=\lambda(\bar{q})$; the effective
potential is defined as
\begin{equation}
 V_{\rm{eff}}(\bar{q})\equiv w(\bar{q})+\beta^{-1}\ln\mu(\bar{q})~.
\label{e.Veff}
\end{equation}

In order to determine the functions $w(\bar{q})$ and $\omega^2(\bar{q})$ we
impose the PQSCHA condition, i.e., we require that the potential
$V(q)$ and the trial potential $V_0(q;\bar{q})$, together with their
derivatives up to second order, have equal averages with respect to the
reduced (diagonal) density $\bar\rho_0(q,q;\bar{q})$.
The condition for the first derivatives is overcome by the definition
of the average point, so that we are left with
\begin{eqnarray}
 \big\langle\!\!\big\langle {V(\bar{q}{+}\xi)} \big\rangle\!\!\big\rangle
 &=& \big\langle\!\!\big\langle {V_0(\bar{q}{+}\xi)} \big\rangle\!\!\big\rangle
 \equiv w(\bar{q})+{m\over2}\omega^2(\bar{q})\,\alpha(\bar{q})~,
\nonumber\\
 \big\langle\!\!\big\langle {V''(\bar{q}{+}\xi)} \big\rangle\!\!\big\rangle
 &=& \big\langle\!\!\big\langle {V''_0(\bar{q}{+}\xi)}
 \big\rangle\!\!\big\rangle
 \equiv m\,\omega^2(\bar{q})~.
\label{e.PQSCHA}
\end{eqnarray}
The latter condition must be solved self-consistently with the definition
of $\alpha(\bar{q})$ in terms of $ \omega^2(\bar{q})$, Eq.~(\ref{e.alpha}).
Now we have all the necessary ingredients to explicitly evaluate the
effective potential and all the thermal averages through the classical-like
expression~(\ref{e.avepq}). As in the case of no dissipation, it can be
seen that for the most usual potentials the self-consistent solution for
$\alpha(\bar{q})$ turns out to be always positive, even though
$\omega^2(\bar{q})$ can be negative \cite{GT86prb,FeynmanK86}.
Indeed, what matters is that $\alpha$, Eq.~(\ref{e.alpha}), is a decreasing
function of $\omega^2$, with a divergence to $+\infty$ [which in the
dissipative case happens at $\omega^2\to-\nu_1^2-\nu_1\gamma(\nu_1)$~].

In the case of {\it Ohmic dissipation}, corresponding to
$\gamma(z)=\gamma=$const [i.e., the memory is Markovian,
$\gamma(t)=\gamma\,\delta(t-0)$], one can see that the ($\bar{q}$-dependent)
mean-square momentum $\lambda(\bar{q})$, Eq.~(\ref{e.lambda}), is divergent
in the Ohmic case. Correspondingly, also $\mu(\bar{q})$, Eq.~(\ref{e.mu})
diverges. The physical reason for this, basically related to the
uncertainty principle, is well discussed in Refs.~\cite{Weiss93,CaldeiraL}.
Here we note that the coordinate mean-square fluctuation $\alpha(\bar{q})$
is still well defined, and that the effective potential can be made finite
by subtraction of the infinite but ${\bar{q}}$-independent quantity
$\beta^{-1}\ln\mu_1$, with
$\mu_1=\prod_{n=1}^{\infty}\big[1+\gamma(\nu_n)/\nu_n\big]$. Therefore,
Eq.~(\ref{e.avepq}) is still meaningful in the case of
observables that do not depend on momentum.
\medskip

For a given potential $V(\hat{q})$, it is convenient to devise a
characteristic energy scale $\epsilon$ (e.g., the barrier height for a
double-well potential, the well depth for physical potentials that vanish
at infinity, etc.) and a length scale $\sigma$ (such that variations of $V$
comparable to $\epsilon$ occur on this length scale) and write
$V(\hat{q})={\epsilon}v(\hat{q}/\sigma)$. In this way one better deals
with the dimensionless coordinate $\hat{x}=\hat{q}/\sigma$. If $x_{\rm{m}}$
is the absolute minimum of $v(x)$, the harmonic approximation (HA) of the
system is characterized by the frequency
$\omega_0=\sqrt{\epsilon{v''(x_{\rm{m}})}/m\sigma^2}$; a dimensionless
coupling parameter $g$ for the system can be defined as the ratio between
the HA quantum energy level splitting $\hbar\omega_0$ and the overall
energy scale $\epsilon$,
\begin{equation}
 g={\hbar\omega_0\over\epsilon}
 =\sqrt{\hbar^2 v''(x_{\rm{m}})\over m\epsilon\sigma^2}~.
\label{e.2.g}
\end{equation}
The case of weak (strong) quantum effects occurs when $g$ is small (large)
compared to 1. In the following application we shall make use of the
dimensionless variables only, i.e., energies are given in units of
$\epsilon$, lengths in units of $\sigma$, frequencies in units of
$\omega_0$, and so on; the reduced temperature is $t=1/(\epsilon\beta)$.
\medskip

Let us then consider the double-well quartic potential
\begin{equation}
 v(x)=(1-x^2)^2~.
\end{equation}
It has two degenerate symmetric minima in $x_{\rm{m}}=\pm{1}$, with
$v''(x_{\rm{m}})=8$. From the PQSCHA equations~(\ref{e.PQSCHA}) and the
definitions~(\ref{e.Veff}) and~(\ref{e.alpha}) we obtain
\begin{eqnarray}
 v_{\rm{eff}}(x)&=&(1-x^2)^2-3\alpha^2(x)+t\,\ln\mu(x)~,
\\
 \mu(x)&=&\prod\limits_{n=1}^{\infty}
 {(\pi n)^2+f^2(x)+\pi n\tilde\gamma(n) \over (\pi n)^2}~,
\\
 f^2(x)&=&{g^2\over 8\,t^2}\,[3x^2+3\alpha(x)-1]~,
\label{e.dwf2}
\\
 \alpha(x)&=&{g^2\over 16\,t}\,\sum\limits_{n=1}^{\infty}
 {1\over (\pi n)^2+f^2(x)+\pi n\tilde\gamma(n)}~,
\label{e.dwalpha}
\end{eqnarray}
where
\begin{eqnarray}
 f(x) &=&{\beta\hbar\omega(x)\over2}={g\over2t}{\omega(x)\over\omega_0}~,
\\
 \tilde\gamma(n)&=&{\beta\hbar\gamma(\nu_n)\over2}
 ={g\over2t}{\gamma(\nu_n)\over\omega_0}~,
\end{eqnarray}
and $\nu_n=(2\pi{t}/g)\omega_0\,n$~.
Equations~(\ref{e.dwf2}) and~(\ref{e.dwalpha}) have to be solved
self-consistently. This task is done numerically; exact reference data
can be obtained only for $\gamma=0$ by numerical solution of the
Schr\"odinger equation.

\begin{figure}[hbt]
\centerline{\psfig{bbllx=8mm,bblly=75mm,bburx=187mm,bbury=211mm,%
figure=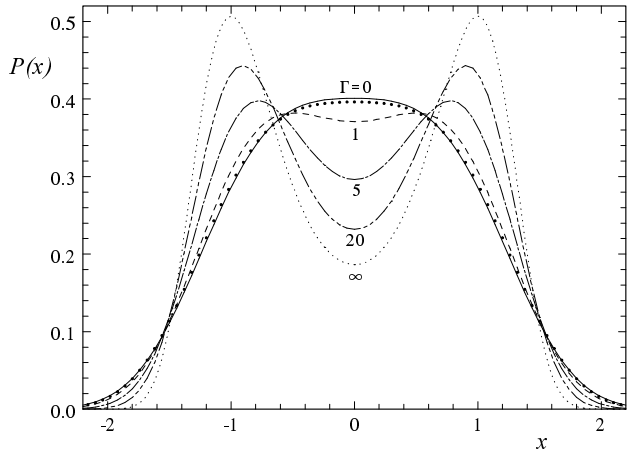,width=82mm,angle=0}}
\caption{Configuration density
${\cal{P}}(x)=\langle\delta(\hat{x}-x)\rangle$ of the double-well quartic
potential for $g=5$, $t=1$, and different values of the Ohmic damping
parameter $\Gamma=\gamma/\omega_0$. The filled circles are the exact result
for $\Gamma=0$; the dotted curve at $\Gamma=\infty$ corresponds to the
classical limit.
\label{f.dwrhot1}
}
\end{figure}

\begin{figure}[hbt]
\centerline{\psfig{bbllx=8mm,bblly=75mm,bburx=187mm,bbury=211mm,%
figure=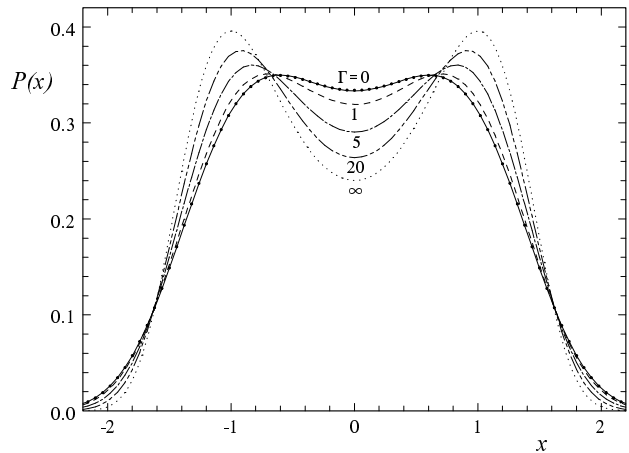,width=82mm,angle=0}}
\caption{Same as figure 1, for $g=5$, $t=2$.
\label{f.dwrhot2}
}
\end{figure}

In Figs.~\ref{f.dwrhot1} and~\ref{f.dwrhot2} we report the shapes of the
coordinate probability distribution
${\cal{P}}(x)=\langle\delta(\hat{x}-x)\rangle$ in the case of Ohmic
damping, $\gamma(\nu_n)=\Gamma\omega_0=\,$const, at a very strong value
of the coupling, $g=5$; this gives a (nondissipative) ground state energy
$e_0=1.394$, and the first excited level is $e_1=e_0+2.355$. When
dissipation is switched on ${\cal{P}}(x)$ tends towards the classical
distribution ${\cal{P}}_{\rm{c}}\sim{e}^{-v(x)/t}$.

\begin{figure}[hbt]
\centerline{\psfig{bbllx=11mm,bblly=75mm,bburx=187mm,bbury=211mm,%
figure=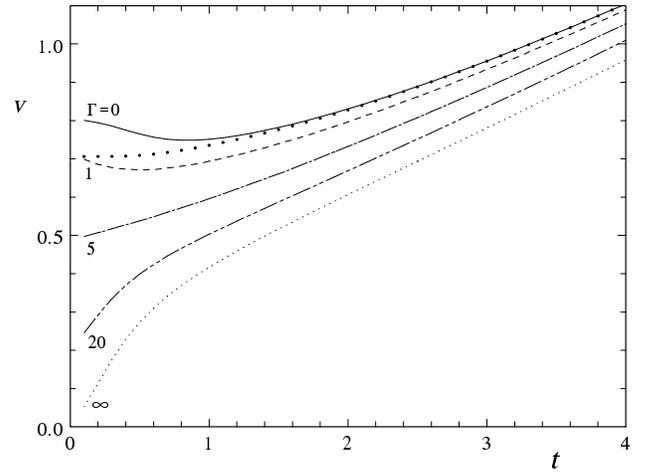,width=82mm,angle=0}}
\caption{Average potential energy $v=\langle{v(x)}\rangle$ of the
double-well quartic potential vs. temperature, for $g=5$ and different
values of $\Gamma=\gamma/\omega_0$.
The filled circles are the exact result for $\Gamma=0$; the dotted curve at
$\Gamma=\infty$ corresponds to the classical limit.
\label{f.dwavevv}
}
\end{figure}

Figure~\ref{f.dwavevv} shows typical results found for the average potential
energy $v(t)=\langle{v(x)}\rangle$. By comparing with the exact data at
$\Gamma=0$ it appears that the PQSCHA gives very accurate results, in spite
of the strong coupling. At lowest temperatures the PQSCHA tends to the
ordinary self-consistent harmonic (or one-loop) approximation, since the
effective Boltzmann factor tends to a $\delta$ function.

However, a more physical model should involve a non-Markovian memory
damping function; in such a model there would be at least one more
characteristic frequency scale (e.g., the frequency $\omega_{\rm{D}}$
in the Drude model \cite{Weiss93}) above which $\gamma(z)$ rapidly vanishes.
In this case all averages make sense, and whereas the averages of
coordinate-dependent observables tend again to the classical behavior (in
other words, the environment quenches the pure-quantum coordinate
fluctuations), those of momentum-dependent ones go in the opposite
direction due to the momentum exchanges with the environment.
Of course, the PQSCHA expressions are even more useful in these physical
situations. Further details and applications, as well as the extension to
the case of many degrees of freedom, will be given in a forthcoming paper.

\medskip
Useful discussions with Professor Ulrich Weiss are acknowledged. We are
also grateful to Dr. Klaus Kirsten (University of Leipzig) for providing us
with an elegant derivation of the tricky expansion of Eq.~(\ref{e.seps}).

\end{document}